\documentstyle[11pt,epsf]{article}

\newcommand{\beq}{\begin{equation}}
\newcommand{\eeq}{\end{equation}}
\newcommand{\beqa}{\begin{eqnarray}}
\newcommand{\eeqa}{\end{eqnarray}}

\begin{document}

\begin{flushright}
MPI-PhT/94-72 \\
hep-th/9411060 \\
November 1994
\end{flushright}

\vfill

\begin{center}{\Large\bf  More on the exponential bound of \\
\smallskip four dimensional simplicial quantum gravity }
\end{center}
\vfill

\begin{center}
\large Bernd Br\"ugmann
\end{center}

\begin{center}
{\em Max-Planck-Institut f\"ur Physik,
 \\ F\"ohringer Ring 6, 80805 M\"unchen, Germany \\}
{\tt bruegman@mppmu.mpg.de}
\end{center}

\smallskip

\begin{center}
\large Enzo Marinari
\end{center}

\begin{center}
  {\em Dipartimento di Fisica and Sezione Infn, Universit\`a di Cagliari,\\
  Via Ospedale 92, 09100 Cagliari, Italy \\}
  {\tt marinari@ca.infn.it}
\end{center}
\vfill

\begin{center}
\large\bf Abstract
\end{center}
\medskip

\noindent
A crucial requirement for the standard interpretation of Monte Carlo
simulations of simplicial quantum gravity is the existence of an
exponential bound that makes the partition function well-defined. We
present numerical data favoring the existence of an exponential bound, and
we argue that the more limited data sets on which recently opposing claims
were based are also consistent with the existence of an exponential bound.

\vspace*{\fill}
\newpage

\section{Introduction}

Theories that we hope could be connected to
Euclidean quantum gravity can be investigated through Monte Carlo
simulations of discrete models, one of the most popular approaches being
dynamical triangulations of simplicial quantum gravity. Here we consider
the four dimensional theory without matter but with a cosmological constant
for fixed topology ${\cal M} = S^4$ \cite{AgMi,AmJu,Br,BrMa}. The most
attractive feature of this model is that there is evidence for a second
order phase transition and therefore a continuum limit may exist.

In this language the Einstein-Hilbert action is very simple. It
depends only on the total number $N_i$ of $i$-simplices contained in
the triangulation $T$:

\beq
  S[T] = k_4 N_4[T] - k_2 N_2[T]\ .
\eeq
The coupling constants $k_4$ and $k_2$ are directly related to the
cosmological constant $\lambda$ and Newton's constant $G$,

\beq
  k_4 = \lambda - \frac{10}{G}\ , \quad k_2 = \frac{2\pi}{\alpha G}\ ,
\eeq
where $\alpha = \mbox{arccos} (\frac14)$. The grand canonical partition
function of the theory is

\beqa
  Z(k_4, k_2) &=& \sum_T e^{-S[T]} \label{Z} \\
  &=& \sum_{N_4} e^{-k_4 N_4} \sum_{T: |T|=N_4} e^{k_2 N_2[T]}\ , \nonumber
\eeqa
where we have split the sum over all triangulations of $\cal M$ into a sum
over all possible volumes (equal to the number of 4-simplices $N_4$) and a
sum over all triangulations $T$ with volume $|T|$ equal to $N_4$, i.e.
the second sum gives the canonical partition function $Z(N_4,k_2)$.

The question that we want to address in this note, and which has been
prompted by \cite{CaKoRe} and further discussed in \cite{BaSm,AmJu2}, is
whether the partition function of such a model is actually well-defined,
i.e. finite.
Suppose that there exists an exponential bound for the canonical partition
function,

\beq
   Z(N_4,k_2) \sim e^{k_4^c(k_2) N_4}\ ,
\eeq

\noindent for large $N_4$ and some constant $k_4^c(k_2)$.
Then the partition function $Z(k_4,k_2)$ is finite for $k_4 > k_4^c(k_2)$
and divergent for $k_4 \le k_4^c(k_2)$.

The question of the existence of an exponential bound for the canonical
partition function is directly related to the asymptotic behavior of the
number of triangulations for a given volume, ${\cal N}(N_4)$, which might
grow as fast as $(5N_4)!$.
Since $2 N_4 < N_2 < 4 N_4$,

\beqa
    {\cal N}(N_4) \le Z(N_4,k_2) \le e^{4 k_2 N_4} {\cal N}(N_4)
    &&	\quad \mbox{if $k_2 \ge 0$,}
\\
    Z(N_4,k_2) < e^{2 k_2 N_4} {\cal N}(N_4) < {\cal N}(N_4)
    &&    \quad \mbox{if $k_2 < 0$.}
\eeqa

\noindent Hence the
existence of an exponential bound on ${\cal N}(N_4)$ implies the same for
the canonical partition function for arbitrary $k_2$, and if there exists
an exponential bound on the canonical partition for a single value $k_2 \ge
0$ then it exists for all $k_2$.

In the next section we discuss how the absence or presence of an
exponential bound manifests itself in the numerical simulations, but first
let us summarize the history of the subject. Until recently the existence of
an exponential bound for the number of triangulations for large volumes was
considered very probable. The bound can be rigorously proven in two
dimensions \cite{TwoDim}. In three dimensions there is strong numerical
evidence for its existence \cite{ThreeDim}, and in four dimensions the
numerical data still seemed reasonably consistent with that claim (e.g.
\cite{AgMi,AmJu,Br,BrMa}).

However recently the authors of \cite{CaKoRe} have claimed that their more
detailed examination of new numerical data shows the absence of an
exponential bound. (In \cite{CaKoRe} the coupling constants are $\kappa_0 =
2 k_2$ and $\kappa_4 = k_4 - 2k_2$.) Their main points are:

\begin{enumerate}

\item
  \protect\label{EN1_EIN}
  In $2d$ everything is fine, and the bound is manifest.
\item
  \protect\label{EN1_ZWEI}
  In $3d$ $k_4^c(N_4)$ has strong power corrections to a (to be proved)
asymptotic constant behavior. One does not see the asymptotic behavior,
but the transient behavior is power like.
\item
  \protect\label{EN1_DREI}
  In $4d$ $k_4^c(N_4)$ is not constant, but changes far less than in $3d$.
The authors of \cite{CaKoRe} assume this means that in this case there is
a logarithmic divergence of $k_4^c$ (while the behavior in $3d$, far more
abrupt, was assumed to be convergent) that implies a violation of the
exponential bound.
\item
  \protect\label{EN1_VIER} The situation in $d\ne 4$ is mainly
  discussed for $k_2=0$, while for $4d$ data is presented also for $k_2 = 0.25$
  and $0.50$, all lower than $k_2^c\approx 1.1$, based on the idea that
  proving existence of the exponential bound at one value of $k_2\ge 0$ is
  sufficient for proving that the partition function is convergent everywhere.

\end{enumerate}

After \cite{CaKoRe}, two papers have discussed the issue in further detail,
both observing that a logarithmic and a small power fit are reasonably
consistent with their data.  The authors of ref.\ \cite{BaSm} lean towards
the logarithmic scaling, and they propose for this case an interesting,
potentially still well-defined scenario, on which we will comment below.
Ref.\ \cite{AmJu2} argues in favor of a power law approach to an exponential
bound, proposing the ansatz of a leading power $\alpha=\frac14$. They fit the
old and some new data that look quite consistent with the ansatz.

Here we present data for $d=4$ and $k_2=0$ for the largest system size yet,
and conclude that the extension to larger systems allows one to decide that
there does exist an exponential bound --- if one is willing to accept as a
compelling evidence the fact that the ratio of $\chi^2$ of the factorial
fit to $\chi^2$ of the exponential fit is $10$. We discuss the strength and
the implications of such a result.

Points (1) through (4) remind us that there is an important issue to be
clarified, but also make the numerically oriented physicist quite
perplex. It is well known indeed that measuring logarithmic corrections is
very difficult (for a classical study see for example \cite{Car}). So in
the case of point (\ref{EN1_ZWEI}) it is impossible to exclude that there
could be a logarithmic divergence, underlying a transient behavior
dominated by power law corrections. The main point in all the following
will be that we are discussing a transient region, with an overlap of
different possible corrections, which could all be present at the same
time.

The same kind of perplexities arise about point (\ref{EN1_DREI}). When
numerical data taken in some limited range of the parameters are
compatible with a logarithmic fit they can also be fitted by a power law
with a small exponent. What one can achieve is putting a bound on the allowed
value of the power, and in our case we will show that indeed a convergent power
dependence has to be preferred over a logarithmic scaling.

We will also discuss in more detail the different $k_2$ regimes (the
crumpled and the elongated phases), but already the evidence contained in
\cite{Br} makes the appreciation of point (\ref{EN1_VIER}) quite delicate.
Indeed, figure $2$ of \cite{Br} (also \cite{AgMi}, and later figure $2$ of
\cite{BaSm}, to be compared) shows that in the two different phases
$k_4^c(N_4)$ is behaving in a very different way. In the elongated phase
$k_4^c(N_4)$ does not depend at all on the volume, while the residual
dependence is all in the crumpled phase. It is possible that because of
strong finite size effects in the crumpled phase, reliable numerical data
about the large volume limit can only be obtained above the apparent phase
transition for $k_2 > k_2^c \approx 1.1 $.

Finally, in \cite{ExBo} hints are given toward the fact that
in $4d$ the number of triangulations could be exponentially bounded.
In three dimensions, there does not yet exist a proof (e.g.\ \cite{DuJo}).
Independent of any analytical arguments, however (and, as far as we know
and understand, we cannot yet rely on a rigorous proof), one should arrive
at the best possible understanding of the numerical data. This point we
will address below, stressing again that we are talking here about the
nature of a transient region, where different corrections of unknown form
may conspire to make the picture difficult to disentangle.

\section{Results for $k_2 = 0$}

When simulating the system described by the partition function $Z(k_4,k_2)$
(\ref{Z}), with variable volume, one finds that there exists a line
$k_4^c(k_2)$ in the plane of the coupling constants such that for any
$k_2$, if $k_4$ is larger than $k_4^c$ the volume tends towards zero
(towards six to be precise), and if $k_4$ is smaller than $k_4^c$ the
volume goes to infinity.  The larger the deviation from $k_4^c$ the faster
is the trend.  For these reasons $k_4^c(k_2)$ is often called the critical
line (although it has nothing to do with statistical criticality
\cite{BrMa}). This critical line at finite volumes represents the
behavior one expects from the discussion of the exponential bound in the
introduction.

A typical Monte Carlo simulation is performed close to a fixed volume $N_4$
and for fixed $k_2$, while $k_4$ is kept close to $k_4^c$ which is determined
dynamically during the simulation. One then looks for a phase transition
determined by some $k_2 = k_2^c$ in the limit of large volume.  How
$k_4^c(k_2)$ is measured is a purely technical question, and the particular
algorithm we use is described in detail in \cite{Br}.

In figure 1 we show $k_4^c$ versus $\ln (N_4)$ for $k_2 = 0$ ($1/G = 0$).
Here we are mainly discussing the point  $k_2 = 0$ because this is where in
\cite{CaKoRe} the strongest argument against an exponential bound is made.
If instead of an exponential bound only a factorial bound holds,
then one expects
\beq
  \protect\label{E_LOG}
  k_4^c(N_4) = a + b \ln N_4\ .
\eeq
If an exponential bound $e^{aN_4}$ to the canonical partition
function exists, then

\beq
  \protect\label{E_POW}
  k_4^c(N_4) = a + b N_4^{-\alpha}\ ,
\eeq
where $N_4^{-\alpha}$ represents a natural polynomial correction to the
exponential. One can argue for $\alpha = 1/4$ \cite{AmJu2}, which allows a
very nice fit to the data. Since there is not enough data to determine
$\alpha$ reliably, setting $\alpha = 1/4$ serves well enough to distinguish
the exponential from the factorial fit.

\begin{figure}
\epsfxsize=400pt\epsffile{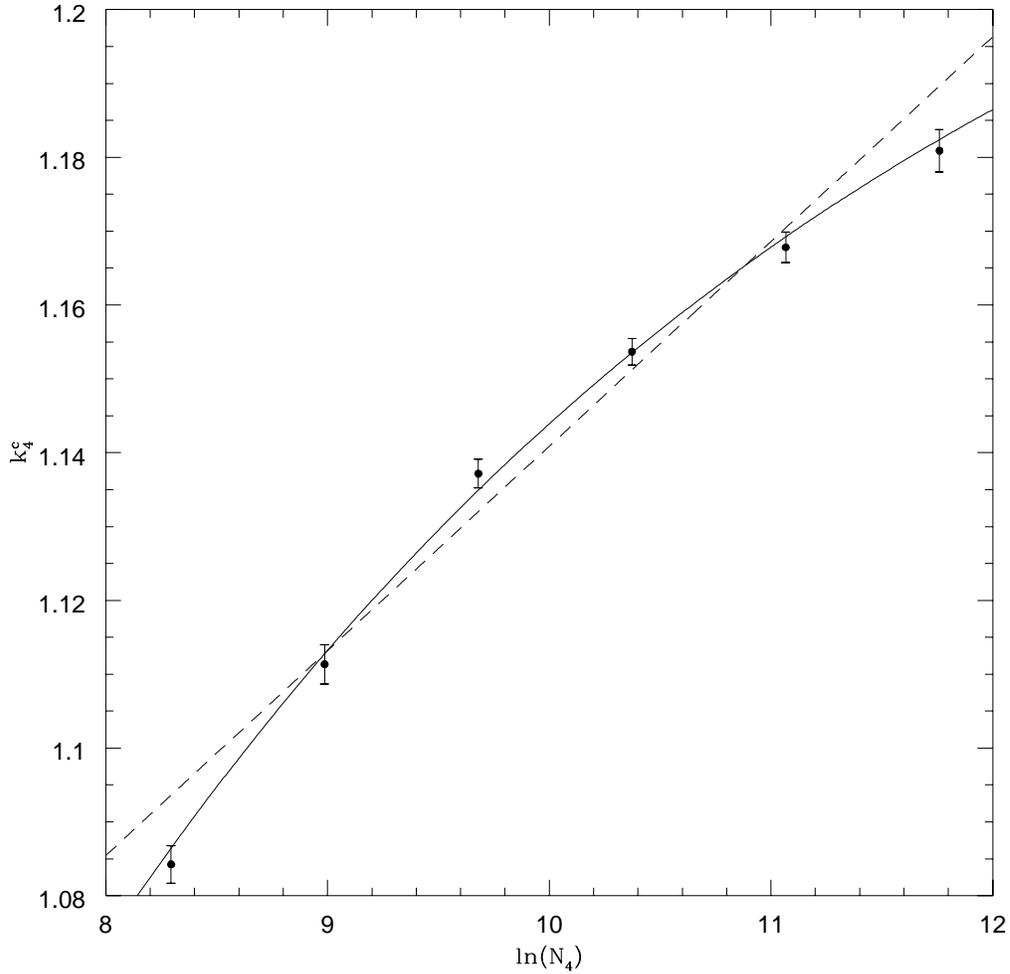}
\caption[a]{\protect\label{FIGURE1}
$k_4^c$ versus $\ln(N_4)$ for $k_2 = 0$. Values for $N_4$ are $4000$,
$8000$, $16000$, $32000$, $64000$ and $128000$.  With the dashed line we
give our best logarithmic fit, with the solid line we give our best fit to
a converging power, with $\alpha=.25$. Both fits have two free
parameters. The $\chi^2$ of the power fit is ten times better than the one
of the logarithmic fit.  }
\end{figure}

In \cite{CaKoRe}, a straight line (we draw our best logarithmic fit with
dashes in figure 1) has been chosen as the best fit to data points between
$N_4$ $=$ $1k$ and $32k$, corresponding to the absence of an exponential
bound due to factorial growth.  In \cite{AmJu2}, a polynomial fit for
$\alpha = 1/4$ is preferred for $N_4$ from 4k to 64k, representing the
existence of an exponential bound which is approached only for still larger
volumes. Volumes of $1k$ and $2k$ 4-simplices have not been included in the
fits since they are likely to suffer from strong finite size effects.
The data of \cite{CaKoRe,BaSm,AmJu2} suggest that we require more data at
larger volumes for a reasonable estimate of the asymptotic behavior.

It is just a matter of computer time to take data at larger volumes ---
more so than a matter of computer memory, and we estimate that our
implementation uses only about one fifth the memory
of the implementation of \cite{AmJu2}.
We have been able to get reliable data up to a volume of $128k$
(50000 sweeps, counting $N_4$ moves actually performed, for $N_4 = 4k$
through $32k$, about 30000 sweeps for $N_4 = 64k$ and $128k$, 10000 sweeps
are discarded for thermalization, the largest volume taking four months on
a (shared) IBM/RISC workstation, autocorrelation time was on the order of
50 sweeps).

It is remarkable that when superimposing our new points
to the fits of ref. \cite{AmJu2} they fall very well on the power fit (quite
far indeed from the logarithmic divergence prediction) obtained from
smaller triangulations.

We have fitted our data for $k_2 = 0$ with the two forms (\ref{E_LOG}) and
(\ref{E_POW}), by setting the power $\alpha=\frac14$. They are both two
parameter fits. Figure $1$ is quite eloquent about the success of the two
fits.  The result is
\beqa
   	k_4^{c(log)} & = & 0.864 + .0277 \ln N_4\ ,
\\
   	k_4^{c(power)} & = & 1.252 - 1.317 N_4^{-\frac14}\ .
\eeqa
The power fit has a value of the $\chi^2 $
which is ten times better than the logarithmic one. We have also
tried $3$ parameter fits. In the power fit we have left the power as a free
parameter, while in the logarithmic fit we have added a volume scale term
$N_4^0$, as in $\ln(N_4-N_4^0)$.  Both fits improve quite a lot, but the
power fit stays far superior to the logarithmic fit (the $\chi^2$ ratio is
now $3$). While such a power fit (where the best power is now $.36 \pm
.04$) matches perfectly the data points, the logarithmic fit is still not
totally congruent to the data (we get $N_4^0$ of order $3000$, that is a
reasonable scale for the transient behavior). We are not very confident in
playing with many parameters, since the allowed corrections are of many
different functional forms, and it is clear that with $6$ data points they
cannot be distinguished. We just take the results of the $3$ parameter fits
as further evidence that the power fit is superior to the logarithmic
fit. Let us also note that indeed the best preferred power is surely not too
small.

The conclusion we draw is that {\it the fits of the numerical data largely
favor the existence of an exponential bound at $k_2 = 0$ over the
presence of a factorial bound.}

\section{Discussion}

What about the consistency of the numerical data? The first observation
about the data in figure 1 should really be that there is a remarkable
agreement in the data from four independent computer implementations
considering that the underlying algorithms are somewhat similar but not
identical. In fact, notice that even the data from \cite{CaKoRe} that lead
to the claim about the absence of an exponential bound curves away from a
straight line in the same way the other data sets do.

Having analyzed in detail the situation for $k_2 = 0$, we now turn to {\it
generic} values of the coupling $k_2$. In theory, the existence of an
exponential bound for any one value of $k_2\ge0$ implies existence for all
the others. But as is well known, but has not been discussed in detail in
this context, there is an important practical difference between the phases
for $k_2$ below and above the critical value $k_2^c \approx 1.1$. For large
positive $k_2$ the simplicial complex is in an elongated phase with an
intrinsic dimension close to two, while for negative $k_2$ the intrinsic
dimension diverges to infinity and the simplicial complex becomes extremely
crumpled. One of the most intriguing and attractive features of simplicial
quantum gravity is that at $k_2^c$ the intrinsic dimension is close to four
\cite{AgMi,BaSm2} (for simplicity we ignore here the problem of giving the
best definition of the intrinsic dimensionality of the system).

The point is that the two phases are not only different, but there is a
genuine asymmetry.  Note that at $k_2^c$ the intrinsic system size for $N_4
= 10,000$ is of the order of $(10,000)^{1/4} = 10$, while at $k_2 = 0$ it
is $(10,000)^{1/10}\approx 2.5$. Therefore, what constitutes a {\em large
volume} that guarantees the absence of finite size effects depends very
sensitively on the value of $k_2$ \cite{Br}. For example, the asymmetry in
the susceptibility present in these systems may be due to such effects.

With regard to the discussion of the exponential bound one should therefore
consider the whole $k_2$ range. Such data already exists in
\cite{AgMi,Br} and were improved upon near the transition in
\cite{BaSm} but were not considered
in \cite{CaKoRe,AmJu2}. For concreteness we show in figure 2 a plot of
$\lambda^c(k_2)$ versus $\lambda_0\sim 1/G$ for $N_4=4k,8k,16k$ based on
\cite{Br}, which for our purpose is better suited than the more accurate
data of \cite{BaSm} since figure 2 extends to extreme values of $k_2$.  The
constants are defined by the relations

\beq
  k_2 = 2 \pi \lambda_0\ ,
  \quad  k_4 = \lambda + 10 \alpha \lambda_0\ .
\eeq

\noindent
There is a definite volume dependence for $k_2 < k_2^c$ while above
the transition no volume effect is discernible.  The linear
transformation from $k_2$ and $k_4$ to the cosmological constant
$\lambda$ is useful for magnifying the volume dependence which is
invisible in this range of coupling constants for $k_4^c(k_2)$
\cite{Br}. This is discussed \cite{BaSm}, but
even when explicitly looking for a small volume dependence for $k_2$
clearly above $k_2^c$, none is found. In this region the plot analogous to
figure 1 appears to be a perfectly horizontal straight line, i.e. there are
no detectable polynomial corrections to the exponential bound.

\begin{figure}
\epsfxsize=350pt
\centerline{\epsffile{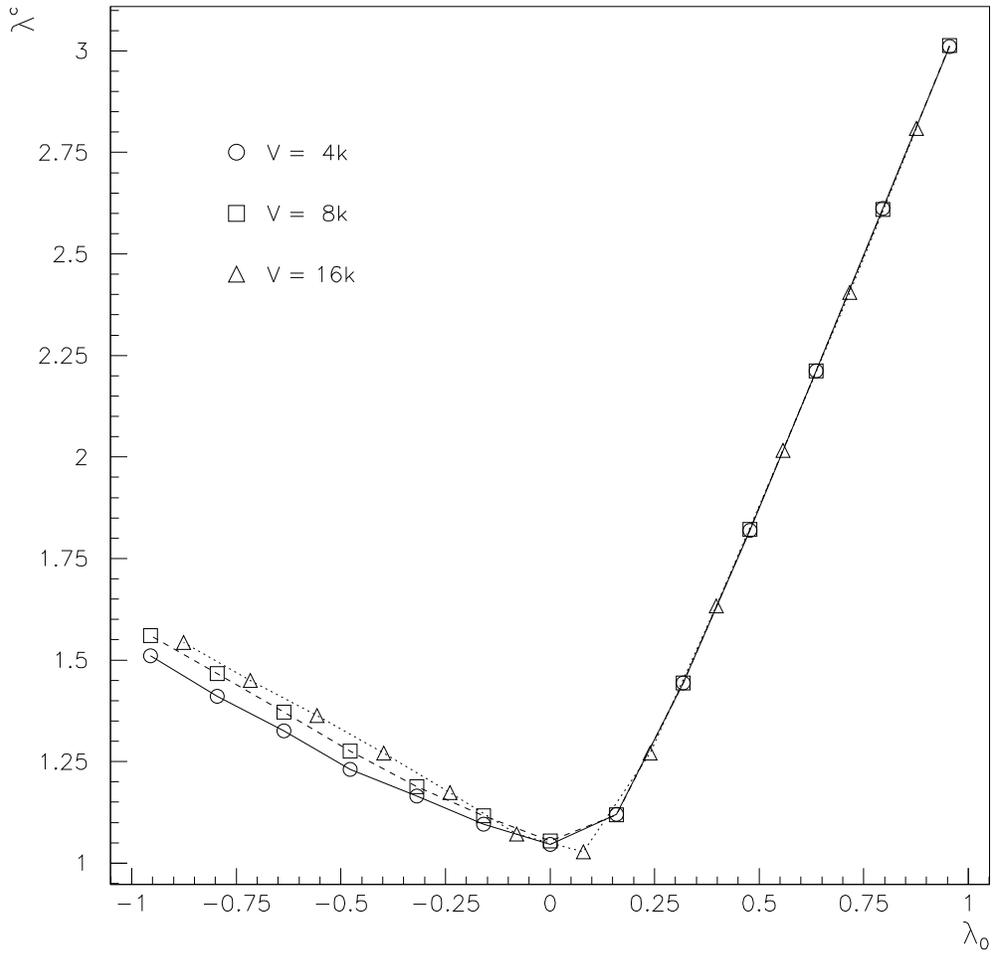}}
\caption[a]{\protect\label{FIGURE2}
$\lambda^c$ versus $\lambda_0$. Indications of a phase transition are
found near $\lambda_0 = 0.18$.}
\end{figure}

The discussion can be taken one step further by noticing that the critical
value $k_2^c$ of $k_2$ moves to larger values with increasing volumes
\cite{AgMi,BaSm}. For larger volumes at $k_2 = 0$ the finite size effects
become even more pronounced (internal dimension up to 50). Given that for
extreme values of $k_2$ the simplicial complex freezes and $k_4^c(k_2)$
becomes a perfect straight line with different slopes, the shift in $k_2^c$
keeping the part $k_2 > k_2^c$ in figure 2 fixed translates directly into
(part of) the volume dependence in the range $k_2 < k_2^c$. When this
effect constitutes the significant part of the volume dependence (for large
enough volume), then the volume dependence of the critical value $k_2^c$
can be estimated by the volume dependence of $k_4^c$ for a small enough but
fixed value of $k_2$. In particular, if there is no exponential bound, then
$k^c_2 \rightarrow \infty$ with $N_4\rightarrow \infty$.

It is instructive to examine the condition for the critical line in the
Monte Carlo simulations (here we follow \cite{Br,BrMa}). Consider the
ergodic random walk in the space of triangulations of $S^4$ consisting of
the five standard moves, where for the move of type $i$ an $i$-simplex is
replaced by a $(4-i)$-simplex. On the critical line, the average volume
$N_4$ is constant, and therefore $N_2$ must also be constant since it is
bounded.  This means that the average variations $\overline{\delta N_j}$
must vanish,
\beq
   \overline{\delta N_j} \sim \sum_{i=0}^4 \Delta N_j(i) p_i = 0,
\eeq
where $p_i$ is the probability with which a move of type $i$ is performed
on average, and $\Delta N_j(i)$ is the change in $N_j$ due to that
move. Since the moves are independent, we obtain
\beq
  p_0 = p_4,  \quad p_1 = p_3,
\label{pppp}
\eeq
on the critical line.

Since the action is linear in $N_2$ and $N_4$, and since the moves are
local, we can be more specific about the conditions on the $p_i$. The $p_i$
can be chosen to be
\beq
	p_i = [e^{-\Delta S(i)}] \, p_i^{geo}\ .
\eeq
The bracket is the Metropolis weight, its key feature being that it depends
only on the type of move and not on the $N_i$ or the triangulation in
general. While the action looks quite trivial, all the non-trivialities
are hidden in the probability $p_i^{geo}$ for a move to be allowed by the
geometric constraints on the triangulation. (Detailed balance is
incorporated in the way the moves are chosen. Potentially, there is a
factor of order $O(1/N_4)$.)

Therefore (\ref{pppp}) is equivalent to
\beqa
	k_4^c &=& \frac{5}{2} k_2 - \ln p_0^{geo},
\label{k1} \\
	k_4^c &=& 2 k_2 - \ln \frac{p_1^{geo}}{p_3^{geo}},
\label{k2}
\eeqa
where we have used that $p_4^{geo} = 1$.
The question of the existence of an exponential bound has therefore been
translated into the question whether there exist appropriate bounds on
the $p_i\equiv p_i(k_4,k_2,N_4)$ which are independent of $N_4$.

First of all, $p_0 \le 1$ implies that $k_4^c$ is bounded from below by
$2.5 k_2$. The hard part is to find a suitable lower bound on $p_0$, for
example, and although it may be possible to do so by some more detailed
analysis of the space of triangulations, we do not have a conclusive
argument. Notice that since moves of type 4 are always allowed, we have
that $p_0 > 0$. However, a naive counting of possible moves of type 0 and
4 around a fixed background triangulation gives $p_0 \sim 1/N_4$, which
would be the divergent scenario, but the same kind of counting would also
make 2d divergent. The counting is, of course, difficult because moves of
type 1, 2, and 3 may change the geometric constraints.

Coming from the numerical side, it is quite suggestive that e.g.\
the data for $N_4=4k$ in figure 2 corresponds to $k_4^c(k_2 \ge 4.0) = 2.497
k_2 + c$ and $k_4^c(k_2 \le -4.0) = 2.002k_2 + d$. This means that in the
extreme $k_2$ regions the relevant geometric probabilities must be
independent of $k_2$ (combining (\ref{k1}) with (\ref{k2}) gives a factor
of $\exp (k_2/2)$ for the opposite side).

Considering the general structure of the phase diagram, the volume
dependence can also be understood on the level of the random walk as
follows. It is the moves of type 4 that drive the system into the elongated
phase ($\Delta N_2(4)/\Delta N_4(4) = 2.5$), while moves of type 1 drive to
the crumpled phase ($\Delta N_2(3)/\Delta N_4(3) = 2.0$).  Depending on
$k_2$ the random walk is driven towards one of the bounds in
$2<N_2/N_4<4$. One of the two possible phases, the elongated phase, is
therefore characterized by low order vertices, and the average order does
not depend on $N_4$ since a maximal elongation can be obtained for a rather
small number of simplices. Hence $p_0^{geo}$, which is the ratio of the
number of vertices of minimal order to the number of all vertices, is
expected to be independent of $N_4$ in the elongated phase. On the other
hand, in the crumpled region the average order of vertices is driven
towards large values, and the average order will grow with
$N_4$. Hence $p_0^{geo}$, which is defined by the low order tail of the
vertex order distribution,  goes to zero with $N_4$ in the crumpled phase.
Equation (\ref{k1}) gives the corresponding volume dependence of $k_4^c$.

In conclusion, when looking for evidence for an exponential bound in the
numerical data of simplicial quantum gravity in four dimensions, one should
take the whole range of $k_2$ into account. If one insists on looking in
the crumpled phase at $k_2=0$, the numerical data strongly support the
validity of an exponential bound.

\newcommand{\bib}[1]{\bibitem{#1}}

\newcommand{\apny}[1]{{\em Ann.\ Phys.\ (N.Y.) }{\bf #1}}
\newcommand{\cjm}[1]{{\em Canadian\ J.\ Math.\ }{\bf #1}}
\newcommand{\cmp}[1]{{\em Commun.\ Math.\ Phys.\ }{\bf #1}}
\newcommand{\cqg}[1]{{\em Class.\ Quan.\ Grav.\ }{\bf #1}}
\newcommand{\grg}[1]{{\em Gen.\ Rel.\ Grav.\ }{\bf #1}}
\newcommand{\jgp}[1]{{\em J. Geom.\ Phys.\ }{\bf #1}}
\newcommand{\ijmp}[1]{{\em Int.\ J. Mod.\ Phys.\ }{\bf #1}}
\newcommand{\jmp}[1]{{\em J. Math.\ Phys.\ }{\bf #1}}
\newcommand{\mpl}[1]{{\em Mod.\ Phys.\ Lett.\ }{\bf #1}}
\newcommand{\np}[1]{{\em Nucl.\ Phys.\ }{\bf #1}}
\newcommand{\pl}[1]{{\em Phys.\ Lett.\ }{\bf #1}}
\newcommand{\pr}[1]{{\em Phys.\ Rev.\ }{\bf #1}}
\newcommand{\prl}[1]{{\em Phys.\ Rev.\ Lett.\ }{\bf #1}}

\end{document}